\documentclass[
     reprint,
     superscriptaddress,
     amsmath,amssymb,amsfonts
     aps,
     prl,
     floatfix,
    ]{revtex4-1}

\usepackage{xfrac}
\usepackage{braket}
\usepackage{graphicx} \graphicspath{{PIC}}
\usepackage{verbatim}
\usepackage{etoolbox}
\usepackage[utf8]{inputenc}
\usepackage{mathtools}
\usepackage{soul,xcolor,colortbl}
\usepackage{hyperref} \hypersetup{colorlinks=true,citecolor=blue,linkcolor=blue,urlcolor=blue}
\usepackage{tabularx}
\usepackage{csquotes}
\usepackage{multirow}
\usepackage{longtable}
\usepackage{cellspace}
\usepackage{resizegather}
\usepackage{dcolumn}
\usepackage[normalem]{ulem}
\usepackage{siunitx}
\usepackage{pdfpages}

\usepackage{bm}
\usepackage{array}
\newcolumntype{C}[1]{>{\centering\arraybackslash}p{#1}}\usepackage{soul}
\definecolor{Gray}{gray}{0.85}

\usepackage{color}
\usepackage{morefloats}

\definecolor{Gray}{gray}{0.9}
\definecolor{LightCyan}{rgb}{0.88,1,1}
\definecolor{green}{rgb}{0.5451,0.2706,0.0745}

\def\CCE{{\small  CCE}}

\def\AFLOW{{\small AFLOW}}

\makeatletter
\AtBeginDocument{\let\LS@rot\@undefined}
\makeatother

\begin{document}

\title{Magnetic State Control of Non-van der Waals 2D Materials by Hydrogenation}

\author{Tom Barnowsky}
\affiliation{Theoretical Chemistry, Technische Universität Dresden, 01062 Dresden, Germany}
\affiliation{Institute of Ion Beam Physics and Materials Research, Helmholtz-Zentrum Dresden-Rossendorf, 01328 Dresden, Germany}
\author{Stefano Curtarolo}
\affiliation{Center for Autonomous Materials Design, Duke University, Durham, NC 27708, USA}
\affiliation{Materials Science, Electrical Engineering, and Physics, Duke University, Durham, NC 27708, USA}
\author{Arkady V. Krasheninnikov}
\affiliation{Institute of Ion Beam Physics and Materials Research, Helmholtz-Zentrum Dresden-Rossendorf, 01328 Dresden, Germany}
\affiliation{Department of Applied Physics, Aalto University, Aalto 00076, Finland}
\author{Thomas Heine}
\affiliation{Theoretical Chemistry, Technische Universität Dresden, 01062 Dresden, Germany}
\affiliation{Institute of Resource Ecology, Helmholtz-Zentrum Dresden-Rossendorf, 04318 Leipzig, Germany}
\author{Rico Friedrich}
\email[]{r.friedrich@hzdr.de}
\affiliation{Theoretical Chemistry, Technische Universität Dresden, 01062 Dresden, Germany}
\affiliation{Institute of Ion Beam Physics and Materials Research, Helmholtz-Zentrum Dresden-Rossendorf, 01328 Dresden, Germany}
\affiliation{Center for Autonomous Materials Design, Duke University, Durham, NC 27708, USA}

\date{\today}

\begin{abstract}

Controlling the magnetic state of two-dimensional (2D) materials is crucial for spintronic applications.
By employing data-mining and autonomous density functional theory calculations, we demonstrate the switching of magnetic properties of 2D non-van der Waals materials upon hydrogen passivation.
The magnetic configurations are tuned to states with flipped and enhanced moments.
For 2D CdTiO$_3$ --- a nonmagnetic compound in the pristine case --- we observe an onset of ferromagnetism upon hydrogenation.
Further investigation of the magnetization density of the pristine and passivated systems provides a detailed analysis of modified local spin symmetries and the emergence of ferromagnetism.
Our results indicate that selective surface passivation is a powerful tool for tailoring magnetic properties of nanomaterials such as non-vdW 2D compounds.

\vspace{0.2cm}

\end{abstract}

\maketitle

\noindent

Two-dimensional (2D) materials --- traditionally deduced from bulk layered compounds bonded by weak van der Waals (vdW) forces --- are emerging as an appealing platform to study fundamental magnetic interactions as well as for spintronic applications.
Ferromagnets were identified
for single layers of Cr$_2$Ge$_2$Te$_6$ \cite{Gong_Nature_2017} and CrI$_3$ \cite{Huang_Nature_2017} with Curie temperatures up to 45~K
beating the Mermin-Wagner theorem due to anisotropy
\cite{Lee_PRL_2020}.
Magnetic phase transitions above room temperature were observed for monolayers of
VSe$_2$ \cite{Bonilla_NNANO_2018} and MnSe$_2$ \cite{Ohara_NanoLett_2018}.

Control over the magnetic properties has been demonstrated via different routes utilizing for instance pressure \cite{Li_NMAT_2019,Webster_PRB_2018}, electrostatic doping \cite{Jiang_NNANO_2018}, or electric fields \cite{Huang_NNANO_2018,Jiang_NMAT_2018,Deng_Nature_2018}.
These magnetic 2D materials are thus promising for spintronics since they can be naturally incorporated into planar device geometries which already led to first demonstration of large magnetoresistance ratios \cite{Klein_Science_2018,Song_Science_2018,Wang_NCOM_2018}.
For such traditional 2D systems,
achieving direct control over the magnetic state via surface functionalization is, however, challenging due to their high chemical stability.

The novel class of non-vdW 2D materials \cite{Balan_MatTod_2022,Kaur_AdvMat_2022} derived from non-layered crystals with strong ionic or covalent bonds offers qualitatively new opportunities based on the engineering of their reactive surfaces.
Several magnetic representatives such as 2D hematene \cite{Puthirath_Balan_NNANO_2018} and ilmenene \cite{Puthirath_Balan_CoM_2018} are derived from natural ores of transition metal oxides hematite ($\alpha$-Fe$_2$O$_3$) and ilmenite (FeTiO$_3$).
These are complemented by other chalcogenides derived from \emph{e.g.} pyrite
(FeS$_2$) \cite{Kaur_ACSNano_2020,Puthirath_JPCC_2021}
and several dozens of systems
suggested by recent data-driven investigations \cite{Friedrich_NanoLett_2022,Barnowsky_AdvElMats_2023,Bagheri_JPCL_2023}.
While these materials are also studied for their catalytic \cite{Puthirath_Balan_NNANO_2018,Puthirath_Balan_CoM_2018}, electrochemical \cite{Kaur_ACSNano_2020}, and optoelectronic \cite{Toksumakov_NPJ2DM_2022} properties, the magnetic features related to potential room temperature ferromagnetism and spin canting are attracting particular attention \cite{Puthirath_Balan_NNANO_2018,Puthirath_Balan_CoM_2018,Padilha_JPCC_2019,Bandyopadhyay_NanoLett_2019,Mohapatra_APL_2021}.

These 2D sheets have been outlined to exhibit (magnetic) cations at their surface \cite{Puthirath_Balan_NNANO_2018,Puthirath_Balan_CoM_2018,Friedrich_NanoLett_2022}
making them appealingly susceptible to environmental influences to control their magnetic state.
Since broken bonds emerge upon exfoliation of the non-vdW 2D systems from the bulk parent compound, chemical control of the magnetic properties due to surface passivation is a promising approach.
As already speculated by Kaur and Coleman \cite{Kaur_AdvMat_2022}, such a surface passivation could be selectively used to create ``activated'' interfaces with desired functionalities.

\begin{figure*}[htb!]
	\centering
	\includegraphics[width=.9\textwidth]{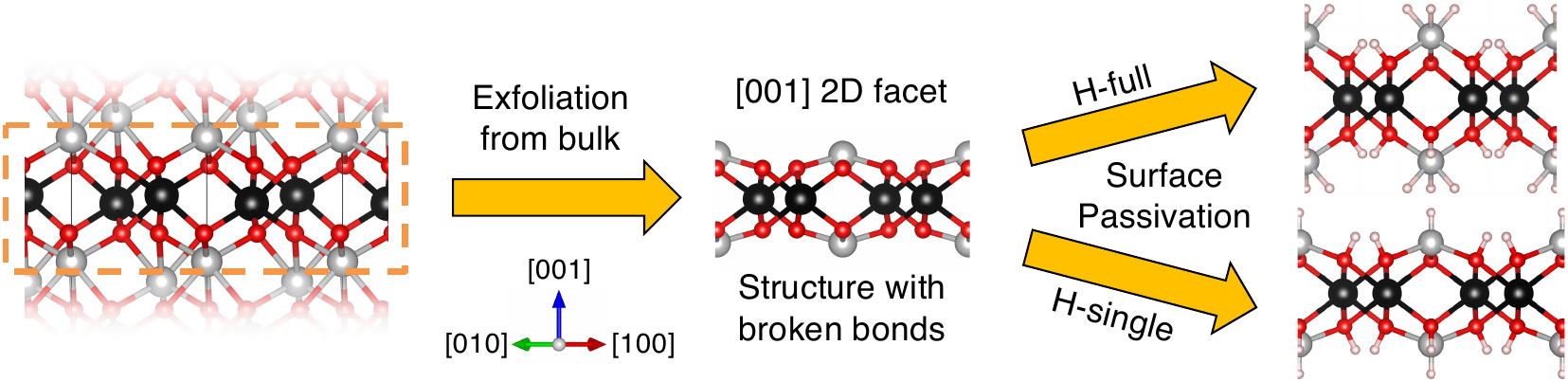}
	\vspace{-0.3cm}\caption{\small \textbf{From exfoliation to passivation.}
	Exfoliation of the 2D facets (orange frame) perpendicular to the [001] direction from the bulk crystal \cite{Puthirath_Balan_NNANO_2018,Puthirath_Balan_CoM_2018,Friedrich_NanoLett_2022} leads to broken bonds to be saturated by two different surface passivation types.
	The compass indicates crystal directions for all structures.
	Colors: cations light gray and black, anion red, H off-white.
	}
	\label{fig0}
\end{figure*}

Here, we show by data mining and autonomous density functional theory calculations, that the magnetic state of non-vdW 2D systems can be controlled by hydrogenation.
Out of an ensemble of pristine candidates, four systems exhibit a net energy gain and
also pass stability tests related to supercell reconstructions and phonon mode analysis.
The magnetic configuration can be switched from ferrimagnetic to antiferromagnetic or to a state with enhanced moments due to hydrogen coverage.
For non-magnetic 2D CdTiO$_3$, the emergence of ferromagnetism upon passivation is observed which can be traced back to a change of the transition metal oxidation states.
The findings are underscored by a visualization of the magnetization density difference
which clearly indicates the breaking of local spin symmetries and the onset of ferromagnetism.

\textbf{Passivation Types:}
We consider the combined set of 35 previously suggested non-vdW 2D materials \cite{Friedrich_NanoLett_2022,Barnowsky_AdvElMats_2023} excluding hematene for which computational surface passivation has already been studied in detail \cite{Wei_JPCC_2020}.
The set contains 32 oxides, two sulfides, and one chloride.
Computational details
are described in section I. in the supporting information ({\small SI}) \cite{PBE,Anisimov_Mott_insulators_PRB1991,Dudarev1998,Oses_CMS_2023,Esters_CMS_2023,kresse_vasp,vasp_prb1996,curtarolo:art104,Esters_NCOM_2021,Friedrich_CCE_2019}.
Here, hydrogen passivation is investigated.
For graphene, it has been reported that reversible hydrogenation can drastically change the electronic properties by transforming the system from the metallic state into insulating graphane \cite{Sofo_PRB_2007,Elias_Science_2009}.
The most intuitive approach is to passivate all (cation-anion) bonds broken upon exfoliation of the slab from the bulk (Fig.~\ref{fig0} upper path).
This corresponds to one H at each of the anions (O$^{2-}$, S$^{2-}$ or Cl$^{-}$) and three at each surface metal cation \cite{Wei_JPCC_2020}.
This passivation type, referred to as \enquote{H-full}, includes 12 H atoms (6 H$_2$ molecules) per unit cell.

To also account for smaller hydrogen load corresponding to a lower H chemical potential, a structure with single passivated surface metal ions and one hydrogen per anion is considered (Fig.~\ref{fig0} lower path).
This \enquote{H-single} passivation has 8 H atoms (4 H$_2$ molecules) per unit cell
\footnote{Test calculations for other passivation options including for example only covering metal ions or anions with hydrogen are energetically highly unfavored and will thus not be considered.}.

\begin{figure*}[ht!]
	\centering
	\includegraphics[width=\textwidth]{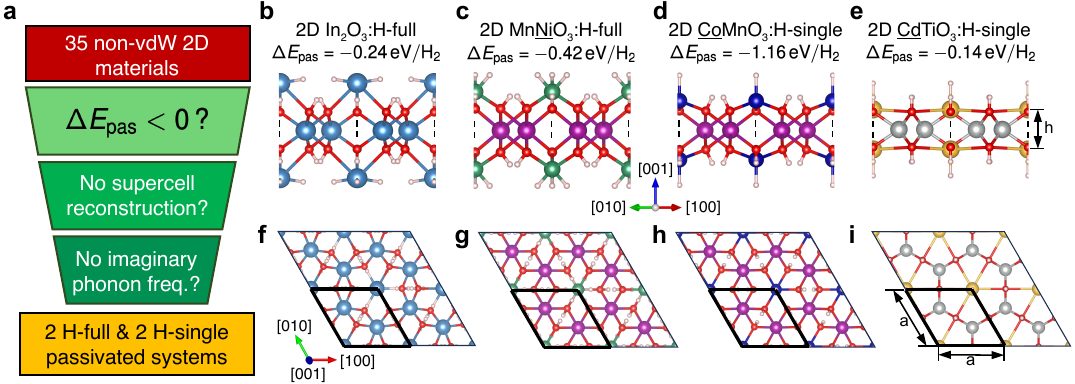}
    \vspace{-0.3cm}
	\caption{\small \textbf{Non-vdW 2D material passivation.}
	(\textbf{a}) Schematic workflow to identify stable passivated non-vdW 2D systems.
	(\textbf{b-e}) Side views of the relaxed structures of passivated candidates, with their respective passivation types and energies indicated.
	The slab thickness $h$, measured between the outer cations, is represented in (\textbf{e}).
	For ternaries, the terminating elements of the pristine slabs (outer ions) are underlined in the formula to clearly indicate the passivated cation species.
	Periodic repetitions of the structure are separated by dashed vertical lines.
	The compass indicates crystal directions for all side views.
	(\textbf{f-i})  Top views of 2$\times$2 supercells of the relaxed structures.
	The in-plane lattice constant $a$ is indicated in (\textbf{i}).
	The area in the black dashed frames corresponds to the respective primitive unit cell.
	The compass indicates crystal directions for all top views.
	Colors: H off-white, O red, Ni green, Mn purple, Co dark blue, In light blue, Ti gray, Cd yellow.}
	\label{fig1}
\end{figure*}

\textbf{Filtering for stable systems:}
Fig.~\ref{fig1}(\textbf{a}) shows the filtering scheme to determine energetically favored and dynamically stable passivated non-vdW 2D systems.
The first criterion for favorable hydrogenation is whether the system is lower in energy than its constituents --- the pristine slab and free H$_2$ molecules --- \textit{i.e.} whether the passivation energy $\Delta E_{\mathrm{pas}} < 0$.
This is fulfilled for four H-full (Rh$_2$O$_3$, In$_2$O$_3$, CoMnO$_3$, and MnNiO$_3$) and nine H-single (Rh$_2$O$_3$, In$_2$O$_3$, AgBiO$_3$, CdTiO$_3$, CoMnO$_3$, MnNiO$_3$, CuVO$_3$, SnZnO$_3$, and BiNaO$_3$) candidates.

To exclude systems susceptible to longer-range reconstructions, 2$\times$2 in-plane supercells are generated from the relaxed geometries with the atomic coordinates randomized (Gaussian distribution, standard deviation 50~m\AA) \cite{Larsen_JPhysCM_2017} and reoptimized.
Out of the 14 candidates, eight lower their energy during supercell relaxation relative to the primitive cell.
Only the remaining three H-full (Rh$_2$O$_3$, In$_2$O$_3$, and MnNiO$_3$) and two H-single (CoMnO$_3$ and CdTiO$_3$) passivated systems are further considered.

To verify whether the remaining five structures are indeed local minima in the potential energy landscape, phonon dispersions are computed \cite{Esters_NCOM_2021} (filtering step three).
The two H-full (H-single) passivated candidates In$_2$O$_3$ and MnNiO$_3$ (CoMnO$_3$ and CdTiO$_3$) do not show imaginary frequencies and are thus vibrationally stable.
The phonon dispersions and density of states for all five systems can be found in section II. in the {\small SI}.

Side/top views of the relaxed structures of the final four systems are depicted in Fig.~\ref{fig1}(\textbf{b}-\textbf{e})/(\textbf{f}-\textbf{i}).
The passivation energies are also included and vary between $-0.14$~eV/H$_2$ for CdTiO$_3$:H-single and  $-1.16$~eV/H$_2$ for CoMnO$_3$:H-single.
All values are much larger than the room temperature thermal energy of $\sim$25~meV indicating that these systems remain passivated at ambient conditions.
At the same time, these moderate $\Delta E_{\mathrm{pas}}$ reveal that the hydrogenation can be reversible as found for graphene \cite{Elias_Science_2009} furthering the value of the associated magnetic state control.

We also estimate the energy penalty due to entropy loss $-T\Delta S$ upon hydrogen coverage.
As an upper bound, one can assume that the entropy of free H$_2$ molecules is completely lost upon passivation (neglecting vibrational entropy effects).
If one takes the tabulated room temperature (298.15~K) entropy of H$_2$ gas of $130.680~$J/(K mol) from the NIST-JANAF thermochemical tables \cite{Chase_NIST_JANAF_thermochem_tables_1998}, $-T\Delta S$ amounts to $\sim0.40$~eV/H$_2$ thus overcompensating the passivation energy of CdTiO$_3$:H-single and In$_2$O$_3$:H-full.
When repeating the same estimate at 100~K, only $\sim0.10$~eV/H$_2$ are obtained indicating the feasibility of passivating these systems at diminished temperatures.

\textbf{Passivated Structures:}
The passivation leads to pronounced structural changes relative to the pristine slab.
The thickness $h$, measured between the two outer cations for comparability to the pristine case, is indicated in Fig.~\ref{fig1}(\textbf{e}), whereas the in-plane lattice constant $a$ is visualized in Fig.~\ref{fig1}(\textbf{i}).
The modification with respect to the pristine compounds are summarized in Table~\ref{tab2}.
The relative changes are defined as $\Delta h=\frac{h_{\mathrm{passivated}} - h_{\mathrm{pristine}}}{h_{\mathrm{pristine}}}$ and $\Delta a=\frac{a_{\mathrm{passivated}} - a_{\mathrm{pristine}}}{a_{\mathrm{pristine}}}$ while the absolute change is the difference between the passivated and pristine values.

\begin{table}[ht!]
    \caption{\small \textbf{Structural Changes.} Relative / absolute changes of slab thickness $\Delta h$ and in-plane lattice constant $\Delta a$ from pristine to passivated systems.
    }\label{tab2}
    \begin{tabular}{l|c|c}
        System & $\quad\Delta h\quad$(\%\,/\,\AA)$\quad$ & $\quad\Delta a\quad$(\%\,/\,\AA)$\quad$\\
        \hline
        In$_2$O$_3$:H-full   & +106\,/\,+3.26  & +2\,/\,+0.14  \\
        MnNiO$_3$:H-full   & +53\,/\,+1.61   & +6\,/\,+0.32  \\
        CoMnO$_3$:H-single & +26\,/\,+0.78   & +12\,/\,+0.66 \\
         CdTiO$_3$:H-single & $-$18\,/\,$-$0.59 & +12\,/\,+0.59
    \end{tabular}
\end{table}

In general, the materials exhibit a thickness increase upon passivation which is particularly pronounced for the H-full scenario reaching 106\% for In$_2$O$_3$.
When initially going from the as sliced bulk to the optimized pristine 2D systems, a thickness reduction on the order of 20-40\% relative to the bulk value was observed due to strong surface relaxations \cite{Friedrich_NanoLett_2022,Barnowsky_AdvElMats_2023}.
The hydrogens saturate the surface dangling bonds created upon exfoliation, thereby counteract this thickness reduction, and even overcompensate it for the H-full systems.
CdTiO$_3$:H-single is a special case \footnote{The unique structural changes upon passivation were verified carefully by obtaining the final structure from different starting geometries.}.
The oxygen atoms rearrange more strongly such that they end up in the same in-plane position at the top and bottom sides of the sheet (see Fig.~\ref{fig1}(\textbf{e}) and (\textbf{i})).
To compensate for the stretched Cd-O bonds, the CdH group moves inwards and thus reduces the slab thickness.

All systems exhibit a moderate increase of the in-plane lattice constants $\Delta a$ up to slightly above 10\% for the H-single systems.
This can be attributed to weaker in-plane bonds between the surface cations and O to compensate for the additional out-of-plane bonding with the passivating hydrogen leading to a slight stretching of the cells.

We focus in the following on how the passivation tunes the magnetic properties.
A comparison of the band structures of the pristine and passivated materials for all systems displaying for instance a modification of the band gaps is discussed in section III. in the {\small SI}.

\begin{figure}[ht!]
	\centering
	\includegraphics[width=\columnwidth]{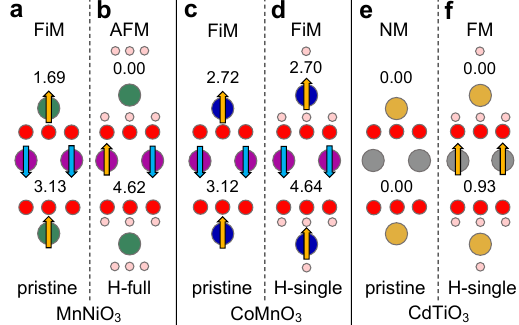}
	\caption{\small
    \textbf{Magnetic configurations.}
    Schematic side views of the magnetic configurations of pristine (\textbf{a}, \textbf{c}, \textbf{e}) and passivated (\textbf{b}, \textbf{d}, \textbf{f}) 2D systems with magnetic moments specified in $\mu_{\mathrm{B}}$ for the selected cations.
    The compositions for which the specific configuration applies are indicated at the bottom.
    }
	\label{fig3}
\end{figure}

\textbf{Magnetism:}
As outlined in Fig.~\ref{fig3}, the magnetic state of three systems critically depends on whether they are passivated or not (In$_2$O$_3$ remains non-magnetic upon hydrogenation).
MnNiO$_3$ displays a switch from a ferrimagnetic (FiM) configuration where the Mn moments of $\sim$3~$\mu_{\mathrm{B}}$ are aligned antiparallel to the surface Ni ones of $\sim$3~$\mu_{\mathrm{B}}$ to an antiferromagnetic (AFM) state with enhanced Mn moments of $\sim$4.6~$\mu_{\mathrm{B}}$ and suppressed spin on Ni.
This behavior can be analyzed by the computed Bader charges.
For the pristine slab, the Mn and Ni charges of 1.88 and 1.21~$e$, respectively, correspond well to the values of 1.89 and 1.21~$e$ for bulk MnO$_2$ and NiO as determined within the \AFLOW-\CCE\ method with similar computational parameters \cite{Friedrich_CCE_2019,Friedrich_PRM_2021}.
In the passivated case, charge is transferred to the Mn and Ni sites leading to 1.48 and 0.66~$e$.
This allows us to identify a switch of the Mn state from +4 to +2 since the value for the passivated system is closer to the 1.37~$e$ known from MnO compared to the $\sim$1.7~$e$ of Mn$_2$O$_3$ from the \AFLOW\ database \cite{Oses_CMS_2023,Esters_CMS_2023}.
This oxidation state fits well to the Mn moment from above close to 5~$\mu_{\mathrm{B}}$ for the $d^{5}$ electron system Mn$^{2+}$.
The strong charge transfer to Ni is indicative of a state close to Ni$^0$.

For CoMnO$_3$, both the pristine and passivated structures show an FiM state albeit with the Mn moment again increased from 3.12 to 4.64~$\mu_{\mathrm{B}}$.
The change of the Bader charges from 1.87 to 1.50~$e$ corroborates again the switch from Mn$^{4+}$ to Mn$^{2+}$.
Also a mild charge transfer to the surface Co$^{2+}$ of $\sim$0.22~$e$ (from $\sim$1.26 to $\sim$1.04~$e$) is found which does, however, not affect its magnetic moment.

Of particular interest is CdTiO$_3$ (Fig.~\ref{fig3}(\textbf{e},\textbf{f})) where an onset of a ferromagnetic (FM) state with Ti moments close to 1~$\mu_{\mathrm{B}}$ from the non-magnetic (NM) pristine sheet is observed \footnote{Several {\small AFM} configurations within a 2$\times$2 in-plane supercell have also been computed and are all higher in energy than the FM state as detailed in section~IV. in the {\small SI}.}.
The change of the Ti charges from $\sim$2.37 to $\sim$2.12~$e$ corresponds well to the data for Ti$^{4+}$ and Ti$^{3+}$ of $\sim$2.40~$e$ and $\sim$2.02~$e$ from bulk TiO$_2$ (rutile) and Ti$_2$O$_3$ \cite{Friedrich_CCE_2019,Friedrich_PRM_2021}.
Like the Ni case above, the strong charge transfer of $\sim$0.8~$e$ (from $\sim$1.20 to $\sim$0.39~$e$) to Cd$^{2+}$ reveals a state close to Cd$^{0}$ upon passivation with no local moment.

Based on the energy difference of $\sim$2.5~meV/formula unit for the FM and AFM magnetic configurations in the primitive unit cell for CdTiO$_3$:H-single, the Ti-Ti magnetic exchange coupling constant $J_{\mathrm{Ti-Ti}}$ and associated Curie temperature can be estimated.
In analogy to Refs.~\cite{Callsen_PRL_2013,Friedrich_PRB_2015,Friedrich_PRBR_2016,Barnowsky_AdvElMats_2023}, the energy difference can be mapped onto a nearest neighbour Heisenberg model $H=-\sum_{\langle i>j\rangle }{J_{ij}\mathbf{m}_i\mathbf{m}_j}$, where $J_{ij}$ is the magnetic exchange coupling constant between ions $i$ and $j$, and $\mathbf{m}_i$ and $\mathbf{m}_j$ stand for the magnetic moments at these sites.
The resulting coupling constant amounts to $J_{\mathrm{Ti-Ti}}\sim 0.82$~meV.
The mean-field Curie temperature $T_{\mathrm{C}}^{\mathrm{MF}}$ can now be estimated from $k_{\mathrm{B}}T_{\mathrm{C}}^{\mathrm{MF}}=J_0/3$ as explained in Ref.~\cite{Lezaic_APL_2007} ($k_{\mathrm{B}}$ is Boltzmann’s constant) by summing all the
coupling constants of one Ti ion as $J_0=3J_{\mathrm{Ti-Ti}}$ \footnote{Our coupling constants are defined as twice the value in Ref.~\cite{Lezaic_APL_2007}}.
We thus arrive at $T_{\mathrm{C}}^{\mathrm{MF}}\sim 10$~K.
However, it is well known that mean-field theory can slightly overestimate Curie temperatures \cite{Lezaic_APL_2007}.

\begin{figure}[ht!]
	\centering
	\includegraphics[width=\columnwidth]{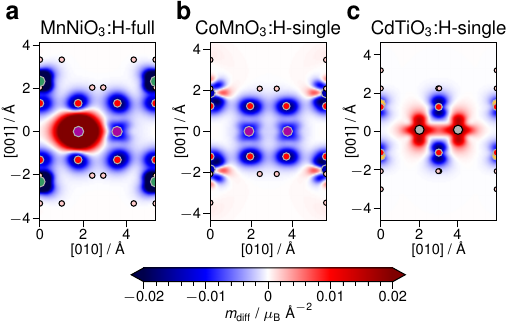}
	\caption{\small
    \textbf{Magnetization density difference.}
    Magnetization density difference (side view) between passivated and pristine MnNiO$_3$ (\textbf{a}), CoMnO$_3$ (\textbf{b}), and CdTiO$_3$ (\textbf{c}).
    Red (blue) indicates a positive (negative) change of the magnetization density due to hydrogenation.
    The small circles represent atoms with the same color code as above.
    Each figure shows one until cell.
    }
	\label{fig4}
\end{figure}

In Fig.~\ref{fig4}, the strong change of the magnetic states upon passivation is further analyzed by magnetization density difference plots (integrated over [100]).
The magnetization density of a system is defined as the difference between spin up and spin down electron densities weighted by the Bohr magneton $\mu_{\mathrm{B}}$: $m(\mathbf{r})=\mu_{\mathrm{B}}(n_{\uparrow}(\mathbf{r})-n_{\downarrow}(\mathbf{r}))$.
As a measure of change, the magnetization density difference between the passivated and pristine system, with the ions in the same positions as in the hydrogenated geometry, can be computed as: $m_{\mathrm{diff}}(\mathbf{r})= m_{\mathrm{passivated}}(\mathbf{r}) - m_{\mathrm{pristine}}^{\mathrm{{\small SCF}}}(\mathbf{r}).$
In practice, the latter quantity is evaluated for a strained pristine structure by a self-consistent field ({\small SCF}) calculation of the passivated geometry with the hydrogens removed.
In the expression, the subtraction of the magnetization density of free hydrogen atoms in the position of the passivated system is omitted.
These have a moment of $1~\mu_{\mathrm{B}}$ each but we are interested in the magnetic transition upon passivation from the pristine to the passivated 2D system which would be heavily obscured by the (artificial) magnetization change at the hydrogens.

For MnNiO$_3$:H-full, Fig.~\ref{fig4}(\textbf{a}) depicts are very strong positive change at the left Mn associated with the flipping and enhancement of the moment at this center upon passivation.
The right Mn only shows a slight negative change due to the increase of the moment without flip.
The data thus signifies the change of the magnetic symmetry between these moments from a mirror relation in the pristine to an inversion in the hydrogenated case.
Also the strong moment reduction around the surface Ni upon passivation is evident from the dark blue region.

The $m_{\mathrm{diff}}$ of CoMnO$_3$:H-single (Fig.~\ref{fig4}(\textbf{b})) reveals the expected slight negative change at the central Mn due to the moment enhancement in the spin down direction.
For the surface Co, both blue and red regions indicate a balance between magnetization gains and losses
fitting well to the above finding of a roughly constant moment.

The dominant feature for the CdTiO$_3$:H-single case (Fig.~\ref{fig4}(\textbf{c})) is the strong magnetization gain at the two central Ti ions in $3d$ orbital shape confirming the onset of the FM state upon passivation.
Note that in all plots, a substantial negative magnetization change at the oxygens is visible.
This is due to the fact that in the strained pristine geometry deduced from the hydrogenated system,
these ions carry a small moment of $\sim 0.3$ to $0.5~\mu_{\mathrm{B}}$ which disappears upon passivation.
The aimed for qualitative analysis of the magnetization change at the transition metal centers is, however, not impacted by this effect.

We have demonstrated control of the magnetic state of transition metal based non-vdW 2D materials by surface hydrogenation utilizing data-mining and autonomous density functional theory calculations.
The modification of the properties was analyzed in detail based on magnetic moments, Bader charges, and magnetization density differences.
For 2D MnNiO$_3$, the hydrogenation induces a transition from an FiM to an AFM state with enhanced Mn moments and a switched local spin symmetry.
In case of 2D CoMnO$_3$, the FiM state is preserved but the Mn moments are again enhanced.
For CdTiO$_3$, the onset of ferromagnetism upon passivation
is revealed.
Our findings thus corroborate that hydrogenation is a versatile and powerful tool to further magnetically ``activate'' these novel non-vdW 2D and likely other nanoscale systems.

\begin{acknowledgments}

The authors thank Mahdi Ghorbani-Asl, Aravind Puthirath Balan, Anand B. Puthirath, Hagen Eckert, Simon Divilov, Corey Oses, Carsten Timm, Thomas Brumme, Agnieszka Kuc, Claudia Backes, and Kornelius Nielsch for fruitful discussions.
R.F. acknowledges funding from the German Research Foundation (DFG), project FR 4545/2-1 and for the “Autonomous Materials Thermodynamics” (AutoMaT) project by Technische Universität Dresden and Helmholtz-Zentrum Dresden-Rossendorf within the DRESDEN-concept alliance.
A.V.K. thanks DFG for the support through Project KR 4866/9-1 and the collaborative research center ``Chemistry of Synthetic Two-Dimensional Materials'' SFB-1415-417590517.
The authors thank the HZDR Computing Center, HLRS Stuttgart (HAWK cluster), the Paderborn Center for Parallel Computing (PC2, Noctua 2 cluster), and TU Dresden ZIH (Taurus cluster) for generous grants of CPU time.

\end{acknowledgments}

\newcommand{\Ozolins}{Ozoli{\c{n}}{\v{s}}}

\clearpage


\section*{Supplementary material}

# Magnetic State Control of Non-van der Waals 2D Materials by Hydrogenation Supporting Information

Tom Barnowsky,[1,2] Stefano Curtarolo,[3,4] Arkady V. Krasheninnikov,[2,5] Thomas Heine,[1,6] and Rico Friedrich[1,2,3,*]

[1]*Theoretical Chemistry, Technische Universität Dresden, 01062 Dresden, Germany*
[2]*Institute of Ion Beam Physics and Materials Research,*
*Helmholtz-Zentrum Dresden-Rossendorf, 01328 Dresden, Germany*
[3]*Center for Autonomous Materials Design, Duke University, Durham, NC 27708, USA*
[4]*Materials Science, Electrical Engineering, and Physics, Duke University, Durham, NC 27708, USA*
[5]*Department of Applied Physics, Aalto University, Aalto 00076, Finland*
[6]*Institute of Resource Ecology, Helmholtz-Zentrum Dresden-Rossendorf, 04318 Leipzig, Germany*
(Dated: October 5, 2023)

## I. Computational Details

The passivation energy $\Delta E_{\text{pas}}$ is computed as the energy difference between the relaxed passivated system $E_{\text{passivated}}$, the pristine non-vdW 2D material $E_{\text{pristine}}$, and free H$_2$ molecules $E_{\text{H}_2}$, normalized per H$_2$:

$$\Delta E_{\text{pas}} = \frac{E_{\text{passivated}} - E_{\text{pristine}} - n\, E_{\text{H}_2}}{n}, \qquad (1)$$

where $n = 4/6$ is the number of hydrogen molecules added to the system for the respective passivation type.

The initial passivated structures are generated from the pristine geometries, with hydrogen bond lengths estimated using the respective covalence radii. A vacuum of at least 20 Å separating periodic images of the slab in [001] direction is inserted before relaxation. Both the ionic positions and the cell shape are allowed to relax, unless stated otherwise.

The calculations for the exchange-correlation functional PBE+$U$ [1–3] are performed with AFLOW [4, 5] and the Vienna *Ab-initio* Simulation Package (VASP) [6, 7] with settings (including DFT+$U$ method and values) according to the AFLOW standard [8] and the internal VASP precision set to ACCURATE. Calculations for slabs are performed on a Γ-centered 10×10×1 $k$-point grid and the phonon dispersions were computed with AFLOW-APL [9] by constructing 3×3 (H-full) or 4×4 (H-single) in-plane supercells (containing 198 and 288 atoms respectively) without polar corrections.

The candidate systems containing potentially magnetic elements such as Ti, V, Cr, Mn, Fe, Co, Ni, and Rh are rigorously checked for magnetism using the algorithm developed within the coordination corrected enthalpies (CCE) method [10], *i.e.* investigating the NM and all possible FM and AFM configurations in the structural unit cell for five different sizes of induced magnetic moments each. The analysis is only applied to other systems when the standard workflow of AFLOW [8] resulted in finite magnetic moments after the relaxation. For CdTiO$_3$:H-single, further longer-range AFM configurations displayed in section III. in the SI are also considered. In each case, the lowest energy magnetic state is used for the further calculations. The reference energy $E_{\text{H}_2}$ was computed in a 10 Å × 10 Å × 10 Å cell using a Γ-only $k$-point sampling while relaxing the H-H bond length.

---

[*] r.friedrich@hzdr.de

## II. Phonon dispersion curves

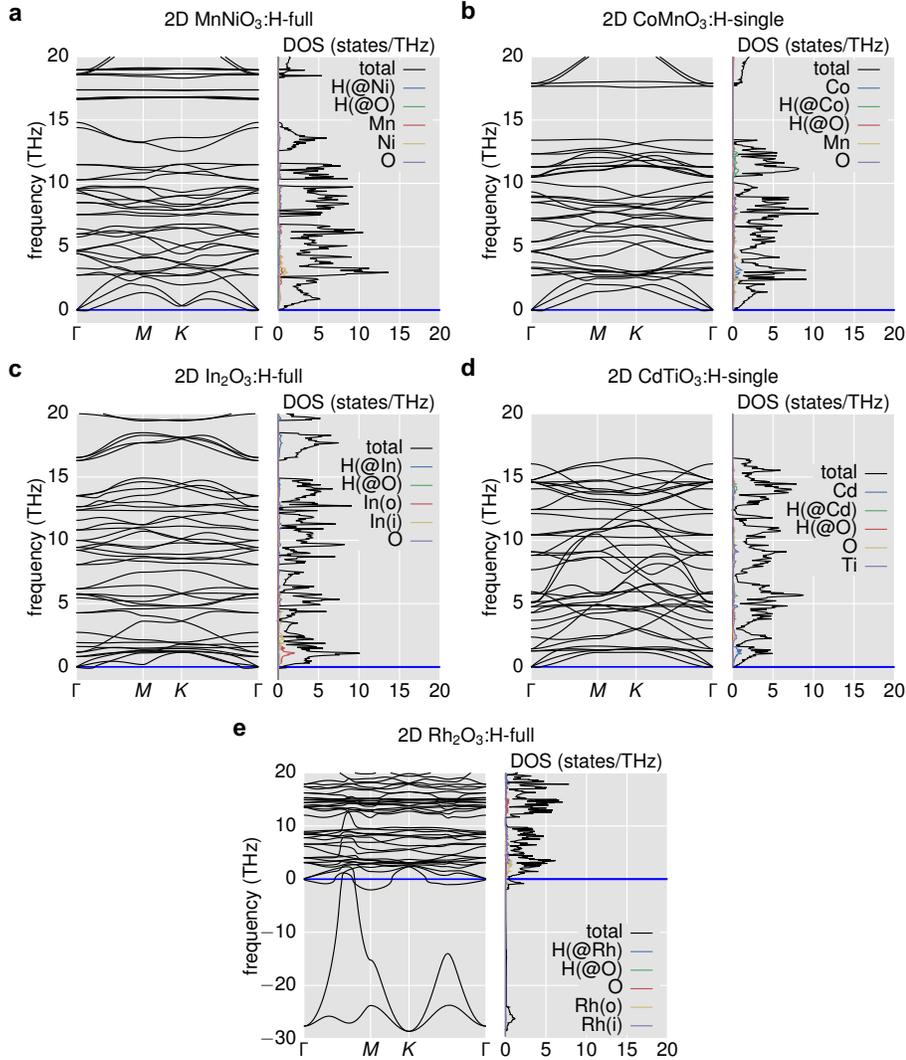

FIG. S1. **Phonon dispersions.** Phonon dispersion curves and density of states (DOS) for the passivated candidates (**a**–**e**) after the supercell reconstruction filter. Contributions by atoms at non-equivalent sites, namely inner and outer cations (labeled (i) and (o) if occupied by the same species for binaries), anions, and hydrogen (labeled (@X) when attached to site X), are indicated in the DOS. Note that a small imaginary dip close to $\Gamma$ is sensitive to numerical details and does not indicate instabilities [11].

## III. Electronic properties

A comparison of the bandstructures of the pristine and passivated 2D systems for each compound is depicted in Figs. S2-S5. For $In_2O_3$/$MnNiO_3$/$CoMnO_3$, the hydrogen coverage leads to a significant opening of the band gaps from $1.29\,\text{eV}/1.79\,\text{eV}/1.29\,\text{eV}$ in the pristine to $2.98\,\text{eV}/2.39\,\text{eV}/2.35\,\text{eV}$ in the passivated case. This opening of the gaps can be understood from the enhanced band splitting upon forming additional bonds to the surface hydrogen and the passivation of surface states originating from exfoliation [12]. $CdTiO_3$ is again a special case since the band gap decreases from $2.75\,\text{eV}$ to $1.96\,\text{eV}$ upon H-single passivation. This can be correlated with the unique structural changes since the strongly elongated bonds in this system give rise to a reduced band splitting.

Also the change in magnetic state is visible in the band structures. While $In_2O_3$ remains non-magnetic when passivated, $MnNiO_3$ switches from ferrimagnetic to antiferromagnetic upon H-full coverage. As such, the band structure of the passivated system depicts no net spin polarization. For $CoMnO_3$, the change of the size of the magnetic moments in the ferrimagnetic state due to H-single passivation goes along a change in the relative spin character of the valence band top and conduction band bottom. Although for the pristine system, the states are both derived from minority spin, for the H-single covered compound, only the valence band top is minority spin derived whereas the conduction band bottom originates from majority spin states. In case of $CdTiO_3$:H-single, the initialization of the ferromagnetism upon passivation causes the states at the gap to both have minority spin character.

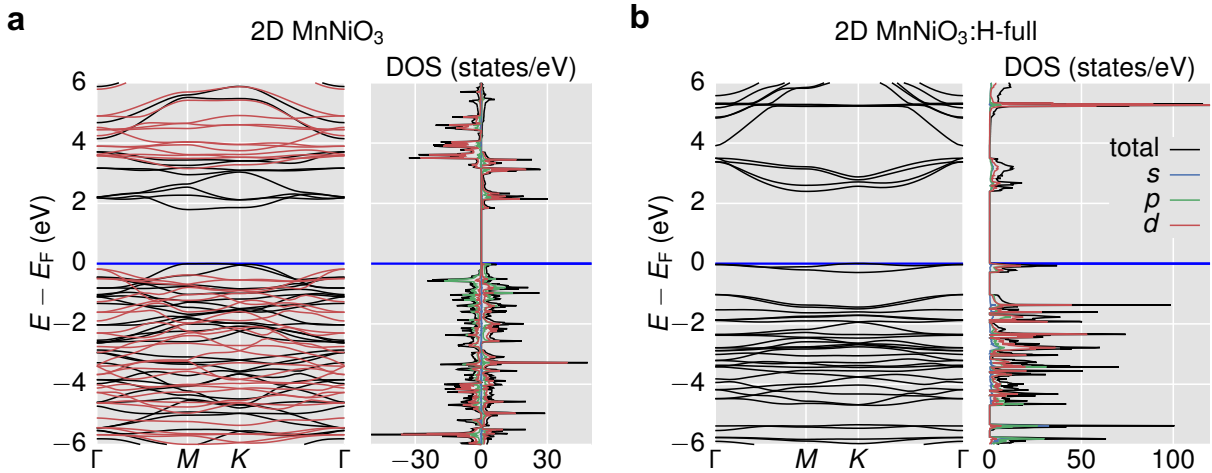

FIG. S2. Bandstructure and density of states for pristine (**a**) and passivated (**b**) $MnNiO_3$. Following the AFLOW standard [8], the energies are aligned at the respective Fermi energy $E_F$ at the top of the valence band. For the spin polarized bandstructure, majority spin bands (positive DOS) are indicated in black, while minority spin bands (negative DOS) are in red.

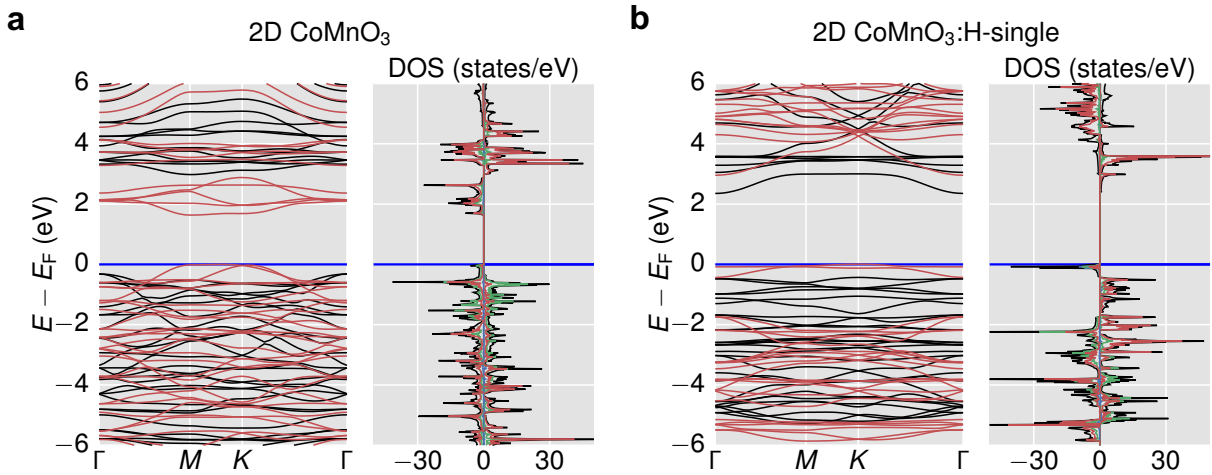

FIG. S3. Bandstructure and density of states for pristine (**a**) and passivated (**b**) $CoMnO_3$.

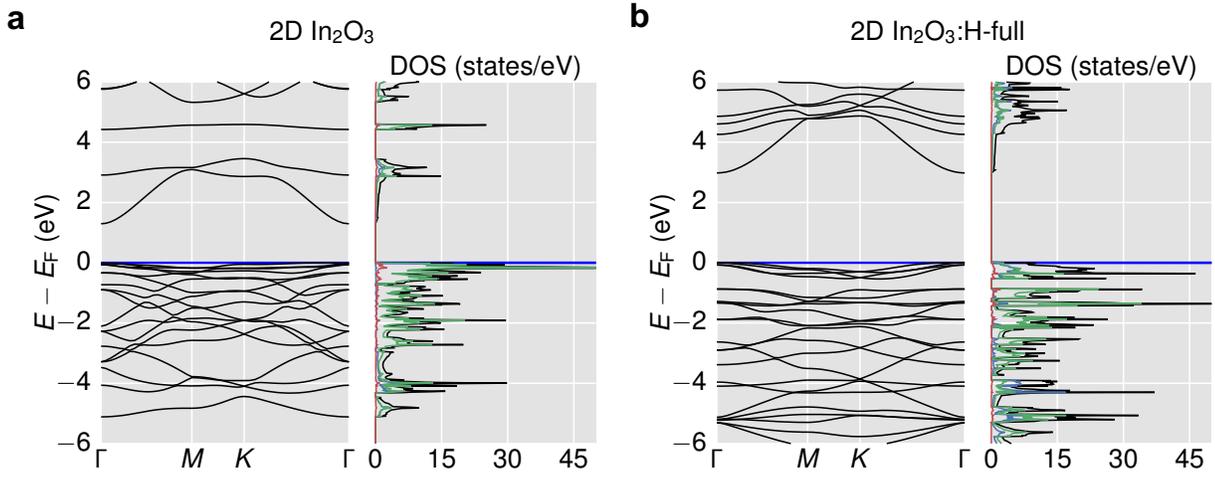

FIG. S4. Bandstructure and density of states for pristine (**a**) and passivated (**b**) $In_2O_3$.

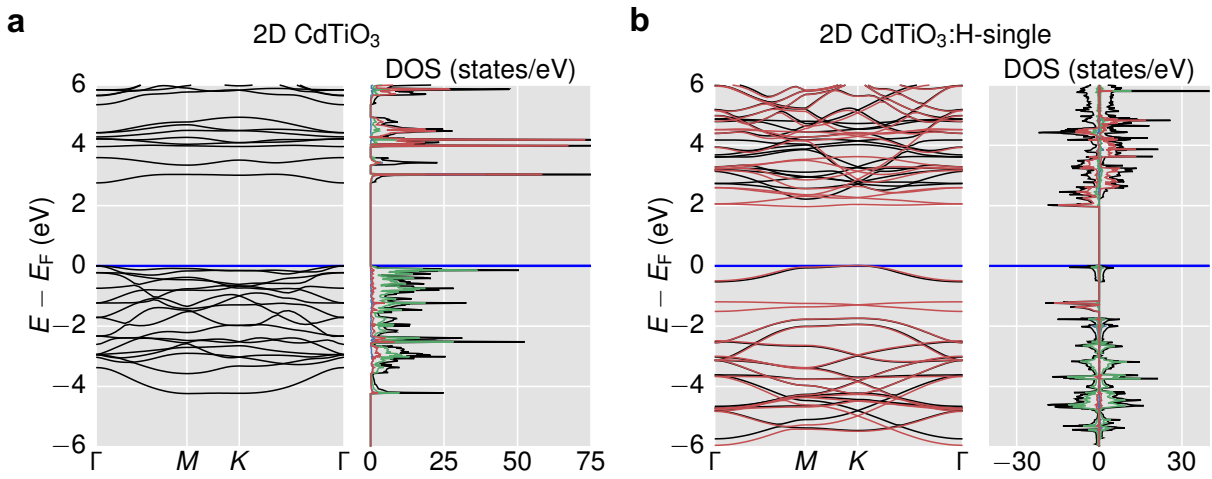

FIG. S5. Bandstructure and density of states for pristine (**a**) and passivated (**b**) $CdTiO_3$.

### IV. Antiferromagnetic configurations for CdTiO$_3$:H-single

As the magnetic unit cell of a system can be in general larger than the structural one, we sampled the magnetic configuration space by four AFM states for CdTiO$_3$:H-single to verify the energetic preference of the FM state. Fig. S6 depicts 2×2 supercells with the respective magnetic configuration of the Ti moments indicated by the color pattern. The first AFM configuration can still be represented in the structural unit cell and has been used to compute the coupling constants but all other states show a longer-range AFM ordering requiring the 2×2 supercell construction. All states are higher in energy than the FM configuration corroborating its energetic favorability.

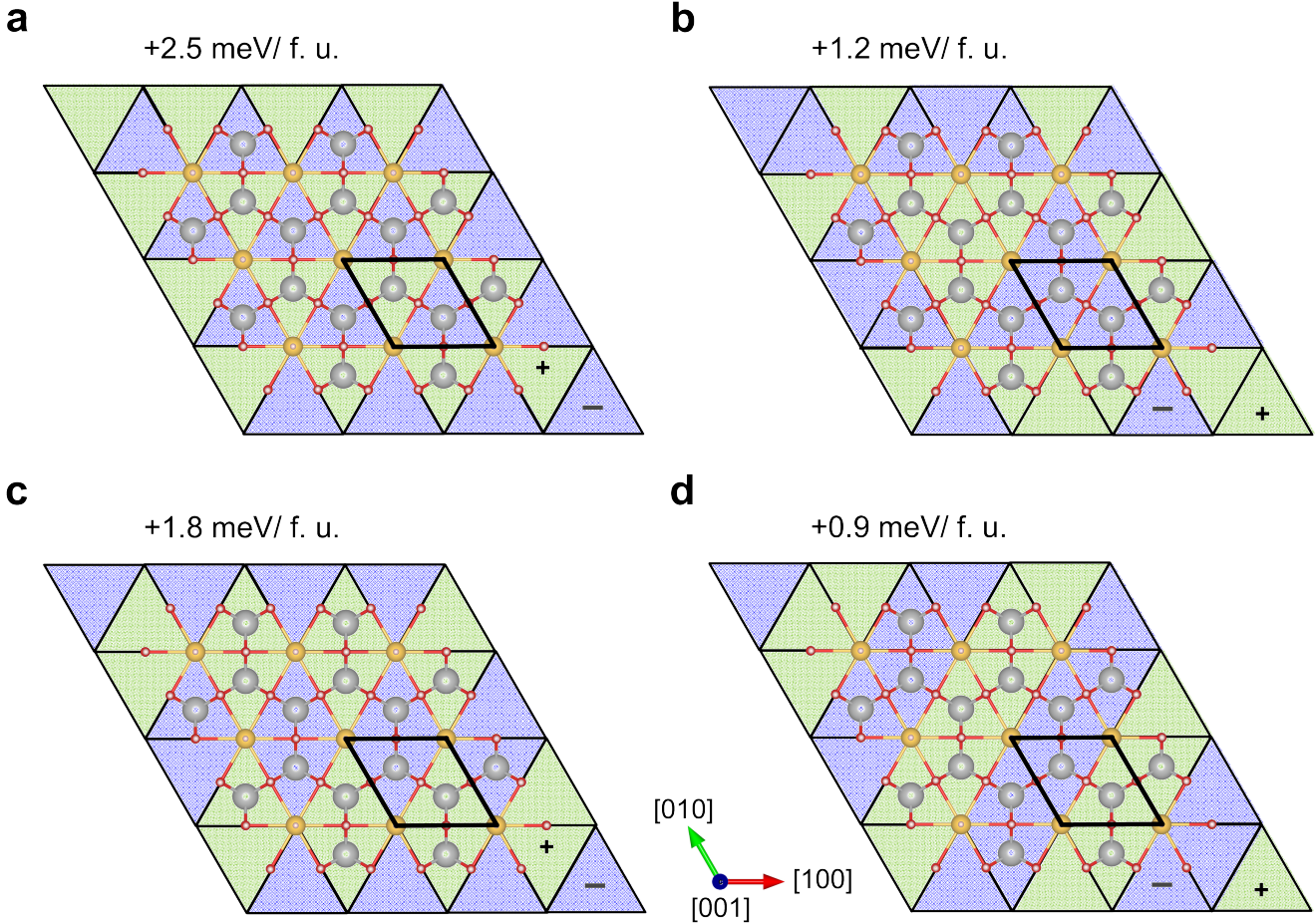

FIG. S6. **Antiferromagnetic configurations for CdTiO$_3$:H-single.** AFM configurations considered for CdTiO$_3$:H-single depicted in a 2 × 2 in-plane supercell. Primitive unit cell (indicated by the black frame) AFM configuration (**a**) and different longer-range AFM configurations (**b** – **d**). The respective energy difference to the FM state in meV/formula unit (f. u.) is indicated in each case. The green (blue) shaded regions around each Ti ion, also extending over the unit cell boundary for better visualization, represent upward,+ (downward,−) alignment of the respective magnetic moments.


\end{document}